# *SEQUENCE RANDOMIZATION USING CONVOLUTIONAL CODES AND PROBABILITY FUNCTIONS*

Vaignana Spoorthy Ella

*Abstract:* This paper investigates the use of different transformations for improving the randomness of sequences. In particular, convolutional codes are used for increasing the size of a given sequence and then a random mapping function is used for further randomization. We have shown how such a method can convert highly correlated sequences into random ones.

## I INTRODUCTION

The randomness of a sequence may be viewed from many different perspectives: physical [1]-[3], algorithmic [4], and probabilistic [5]. The only guaranteed way to ensure perfect randomness is to use a quantum process [6] but even there questions related to initialization complicate the picture [7],[8]. Random sequences have many applications in cryptography and we are interested in a method that can convert a highly correlated sequence to a random one. Statistical tests may be used find out whether the transformed sequence is truly random or not. One of these is the runs test. A run is a series of increasing values or a series of decreasing values of some function of the successive elements of the sequence. In a random sequence, the probability that the *i*th value is larger or smaller than the *(i-1)*th value follows a binomial distribution, and this is the basis of the runs test. The performance of the runs test approximately mirrors two-valued autocorrelation function.

The method described in this paper can begin with a sequence which can have very non-equal numbers of 0s and 1s. We use a probability distribution algorithm to increase the randomness of this starting sequence by converting it into another sequence which has equal number of 0s and 1s. This process also increases the likelihood that the final sequence will pass tests of randomness.

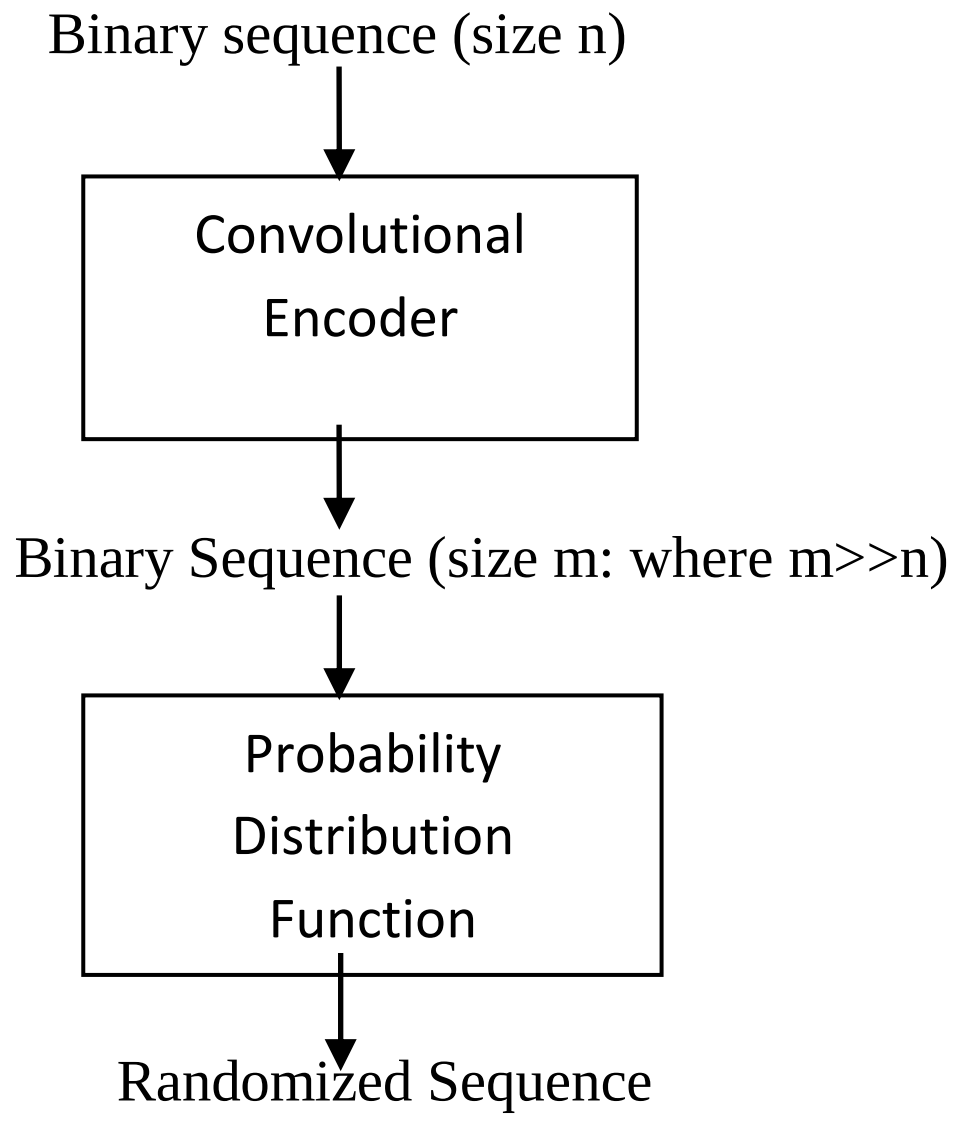

*Figure 1 General architecture of the proposed system*

Our system uses a convolutional coder [9],[10] to increase the size of the given input sequence. Later, we use probability mapping algorithm to improve its randomness.

## II THE PROPOSED SYSTEM

Consider ($m_0$, $m_1$, $m_2$...$m_n$) as the message bit, $V_n$ (n=0, 1 ...) as the output bit in a given step and C as the final output of the Convolutional encoder. $_0$, $G_1$, $G_2$ are the generator polynomials. The generator polynomial gives which message bit should be used to perform XOR operation. In the below example, it is given that $G_0$= (1, 0, 1), it means we should perform XOR operation on message bits $m_0$ and $m_2$ only.

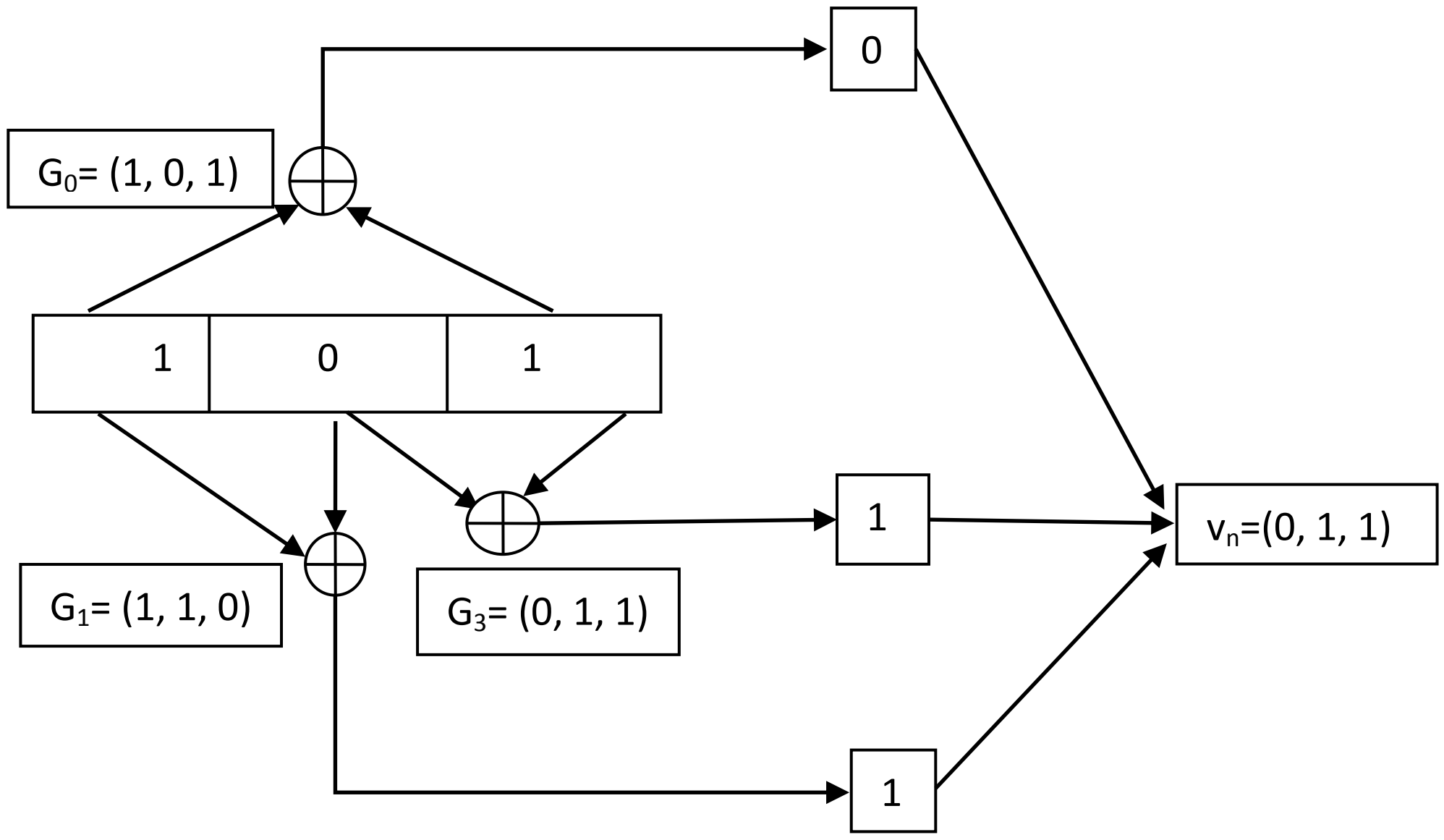

*Figure 2: Example of convolutional Encoder*

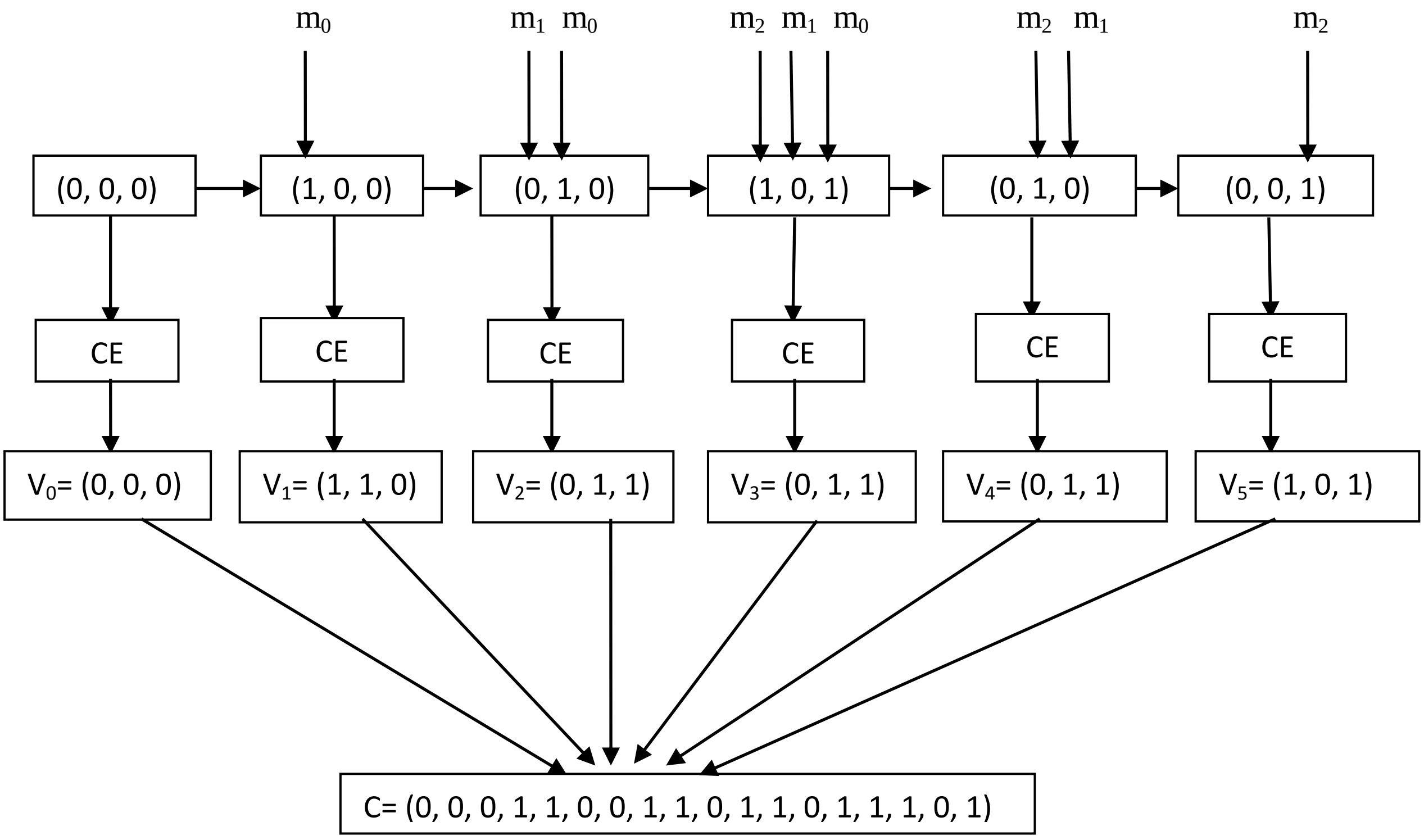

*Figure 3 Right Shift operations and working of Convolutional encoder (CE)*

The input message bit for the convolutional encoder is (1, 0, 1). Convolutional encoding is done by performing right shift operations on the given message bit.

*Step 1:* Convolutional encoding starts with (0, 0, 0) as input message bit, then we get output as $v_0$= (0, 0, 0).

*Step 2:* perform Convolutional encoding by taking (1, 0, 0) as input message bit, then we get output as $v_1$=(1, 1, 0)

*Step 3:* perform Convolutional encoding by taking (0, 1, 0) as input message bit, then we get output as $v_2$=(0, 1, 1)

*Step 4:* perform Convolutional encoding by taking (1, 0, 1) as input message bit, then we get output as $v_3$=(0, 1, 1)

*Step 5:* perform Convolutional encoding by taking (0, 1, 0) as input message bit, then we get output as $v_4$=(0, 1, 1)

*Step 6:* perform Convolutional encoding by taking (0, 0, 1) as input message bit, then we get output as $v_5$=(1, 0, 1)

Now combine all these outputs to get the final output of convolutional encoding.

Finally, we get output sequence as (0, 0, 0, 1, 1, 0, 0, 1, 1, 0, 1, 1, 0, 1, 1, 1, 0, 1) and the corresponding input message sequence is (1, 0, 1). In this way, we can increase the length of the given message sequence using convolutional codes.

However, the output sequence we got from the convolutional encoder is not random. It contains more number of 1s than 0s. Now our goal is to make number of 1s as equal as number of 0s. To achieve this, we make use of the probability distribution function.

## III THE PROBABILITY DISTRIBUTION ALGORITHM

The goal of probability distribution algorithm is to make number of 1s and number of 0s in a given sequence as equal.

*Case 1:* If the number of 1s is more than number of 0s:

It randomly chooses a message bit randomly (Here we make use of java random function), if the message bit is 1, then it converts it into 0 with a probability= ([no. of 1s-no. of 0s]/2)/no. of 1s, and else if the message bit is 0, then it will not alter the bit .This algorithm is repeated until we get equal no. of 0s and 1s in a given sequence.

*Case 2:* If the number of 0s is more than number of 1s:

It randomly chooses a message bit randomly (Here we make use of java random function), if the message bit is 0, then it converts it into 1 with a probability= ([no. of 0s-no. of 1s]/2)/no. of 0s, and else if the message bit is 1, then it will not alter the bit .This algorithm is repeated until we get equal no. of 0s and 1s in a given sequence.

IV IMPROVED JAVA RANDOM FUNCTION

In the existing Java random function, we choose a random number from the given range. Further, in order to improve the randomness of Java random function, a random range from the given range is chosen in order to pick the random number.

**Existing Java Random Function:**

*Random_Number* = Java Random ( int Range);

**Improved Java Random Funtion:**

Random_Range= Java Random (int Range);

*Random_Number* = Java Random (int Random_range);

**Performance Analysis of  Improved Java Random Function:**

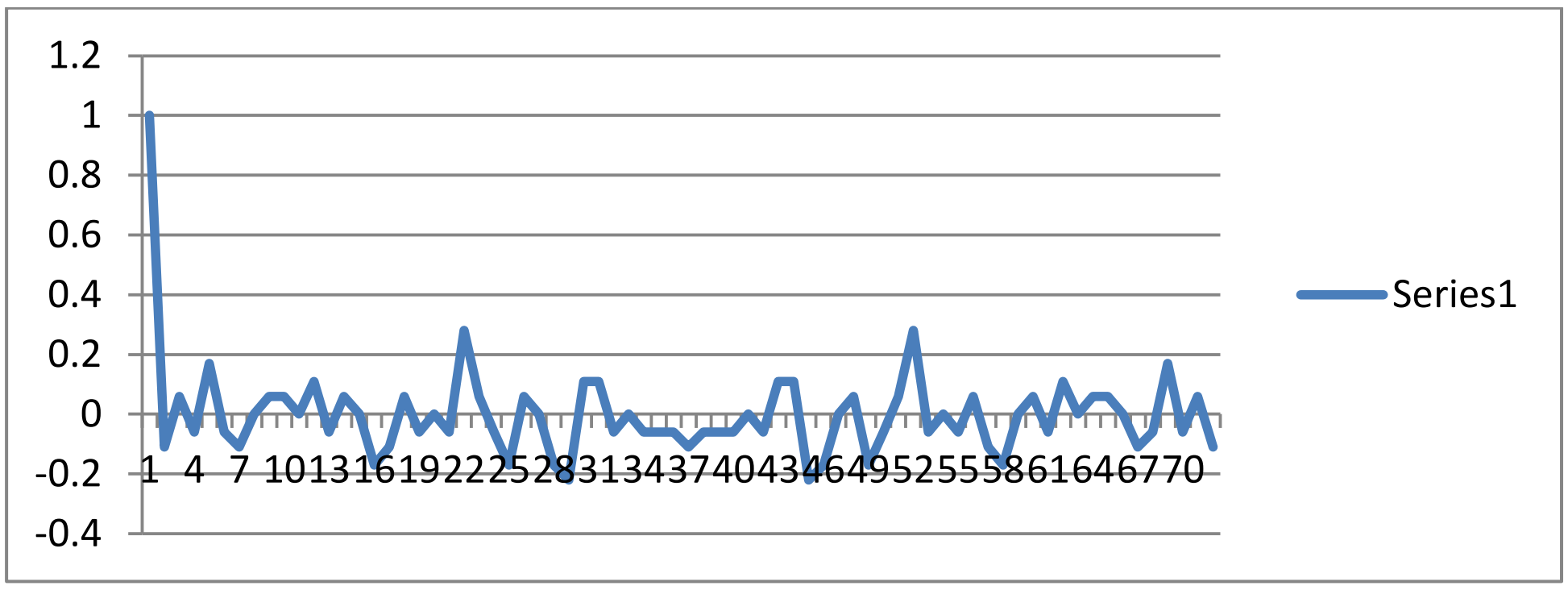

Using Old Java Random function Input:  72 all zeros peak value: 0.28

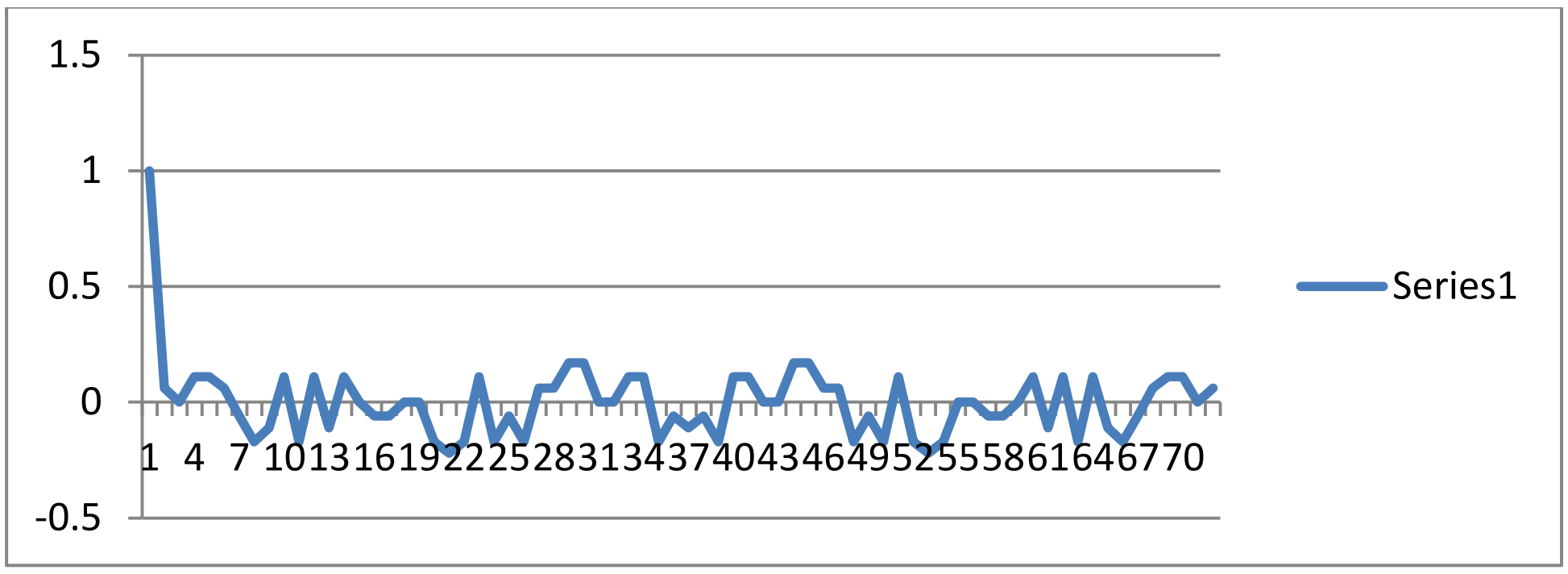

Using New Java Random function Input:  72 all zeros peak value: -0.22

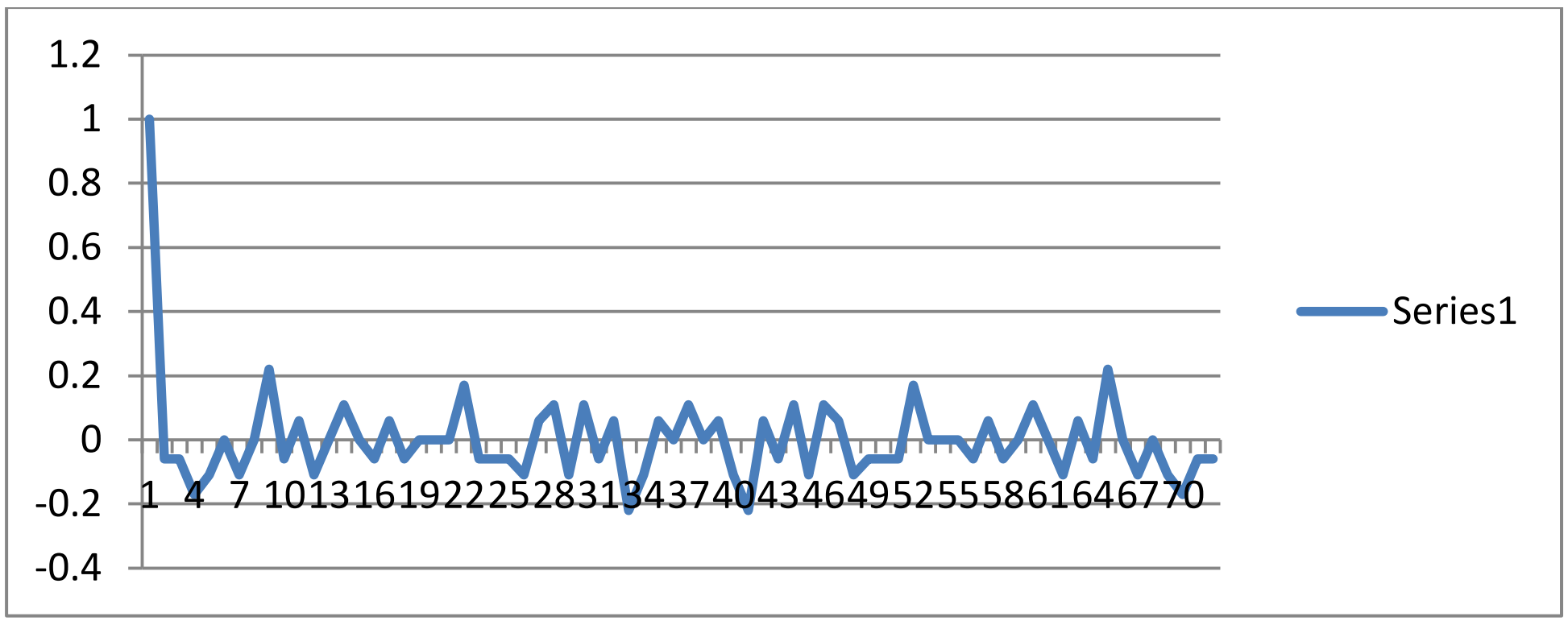

Using Old Java Random function Input:  71 zeros 1 One peak value: -0.22, 0.22

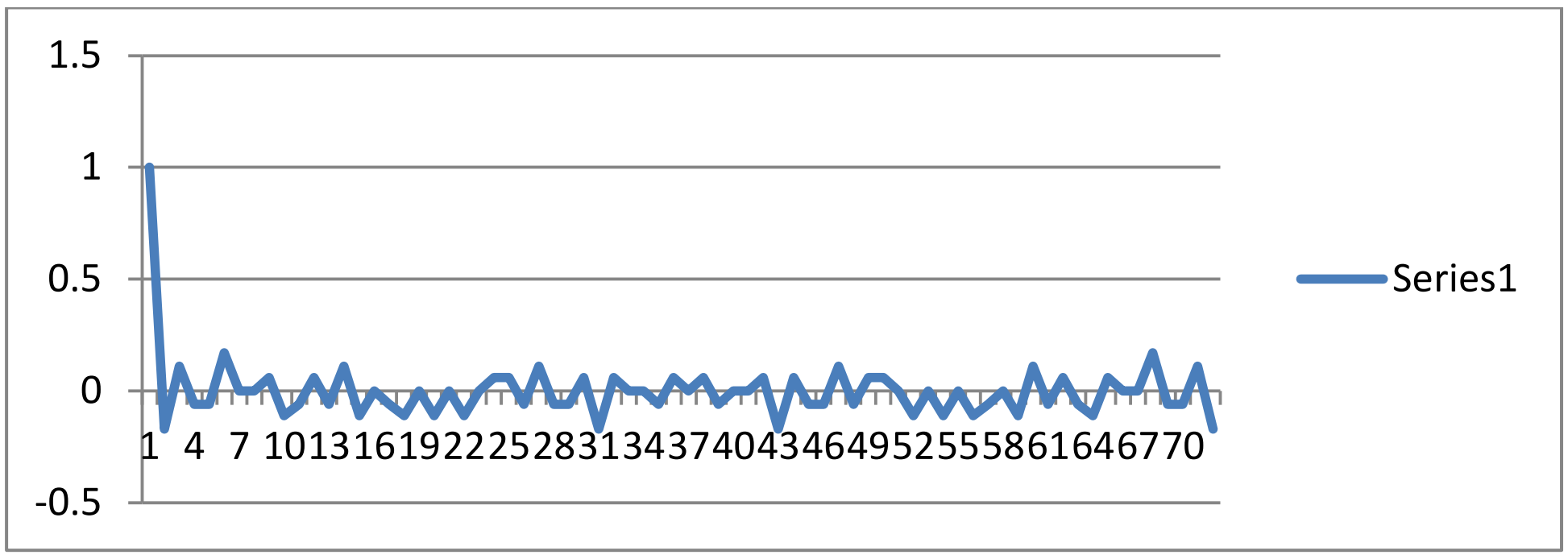

Using New Java Random function Input:  71 zeros 1 One peak value: -0.17, 0.17

Observations: Improved Java random function shows its performance as the length of the sequence increases. For a small sequence there is no much difference between New and Old Java random functions.

| Length of the Sequence | Peak value of Old Java Random Function | Peak value of New Java Random Function |
|---|---|---|
| 10 | 0.09 | 0.09 |
| 220 | 0.08 | 0.06 |
| 550 | 0.11 | .08 |
| 770 | +0.22 , -0.22 | +0.17 , -0.17 |
| 9801 | 0.07 | 0.03 |

Clearly the peak value of autocorrelation for the New Java random function is much reduced compared to the Old Java random function.

## V PROBABILITY DISTRIBUTION ALGORITHM USING IMPROVED JAVA RANDOM FUNCTION

As example consider the input message sequence to be (0, 1, 1, 1, 0, 1, 1, 0, 1, 1). It has number of 1s=7, and number of 0s=3. The probability of 1s=7/10 and probability of 0s= 3/10.

The number of 1s to be converted into 0s= (7-3)/2=4/2=2 . The probability no of 1s to be converted into 0s is (7/10*2/7) =1/5.

Here message length is 10 hence (10*1/5) =2. Hence two 1s should get converted into 0s.

Now randomly choose two random 1s (random indexes of 1s are chosen by Java random function) and replace them with 0s. For example, choose third position and 9th position 1s and convert them into 0s. The output message sequence we get is (0, 1, 0, 1, 0, 1, 1, 0, 0, 1), which has equal number of 0s and 1s.

## VI PERFORMANCE MEASURE OF THE PROPOSED SYSTEM

Here we measure the performance of the proposed system using the autocorrelation function [11]. We use the D-sequences as initial random sequence in several of the tests [12]-[15]. More comprehensive testing can also be done [16] but will not be considered in this paper.

### *1. Tests of Randomness*

We use autocorrelation function to test the randomness of a given sequence. While calculating autocorrelation values 0s are replaced with -1. If the given sequence is random, autocorrelation values should be nearer to zero.

### *2. Autocorrelation*

The autocorrelation function is a measure of the randomness of the sequence. It is given by

$$C(k) = \frac{1}{n}\sum_{j=1}^{n}(a_j a_{j+k})$$

where n is period and k=0 to n-1.

### *3. Randomness Measure*

Randomness R of a sequence of period n is measured by the following formula. If the given sequence is random, R value should be nearer to 1and for a constant sequence the randomness measure is 0.

$$R(Sequence) = 1 - \frac{1}{n-1}\sum_{k=1}^{n-1}(|C(k)|)$$

**Example 1:** *Input sequence: all zeros and last bit is 1.*

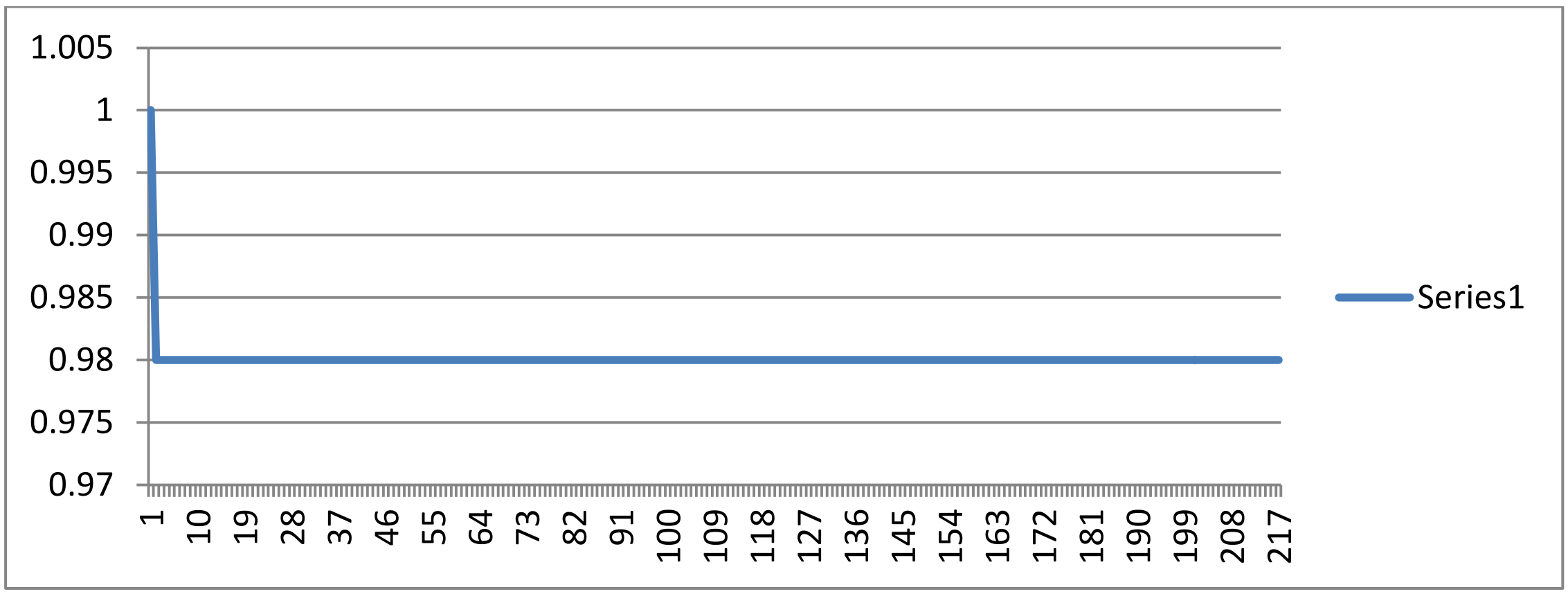

*Figure 4 Autocorrelation graph for input sequence.*

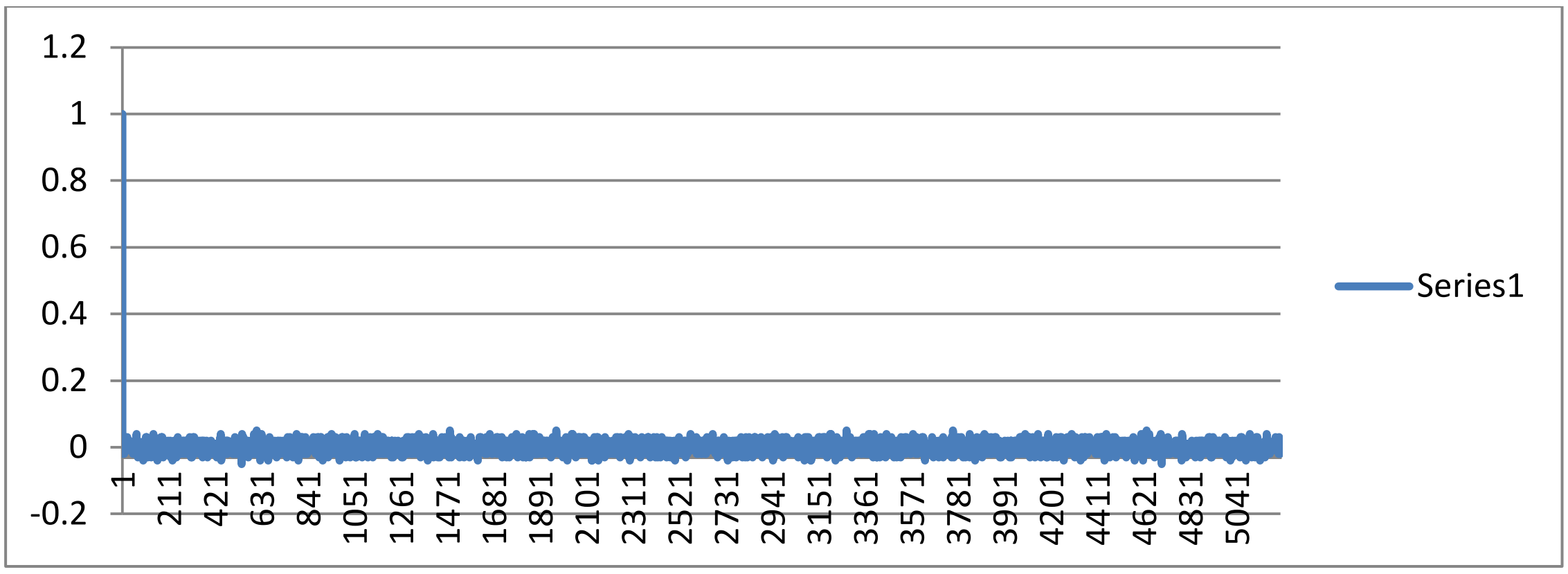

*Figure 5 Autocorrelation graph for output sequence generated by probability function.*

This example shows how a completely deterministic sequence is transformed effectively into a random sequence.

**Example 2:** *Input Sequence type: D-sequence of size 1019 for which the period=1018.* As we know such a sequence has an asymmetry across its mid-point which shows up in the value of -1 in the autocorrelation for half the period.

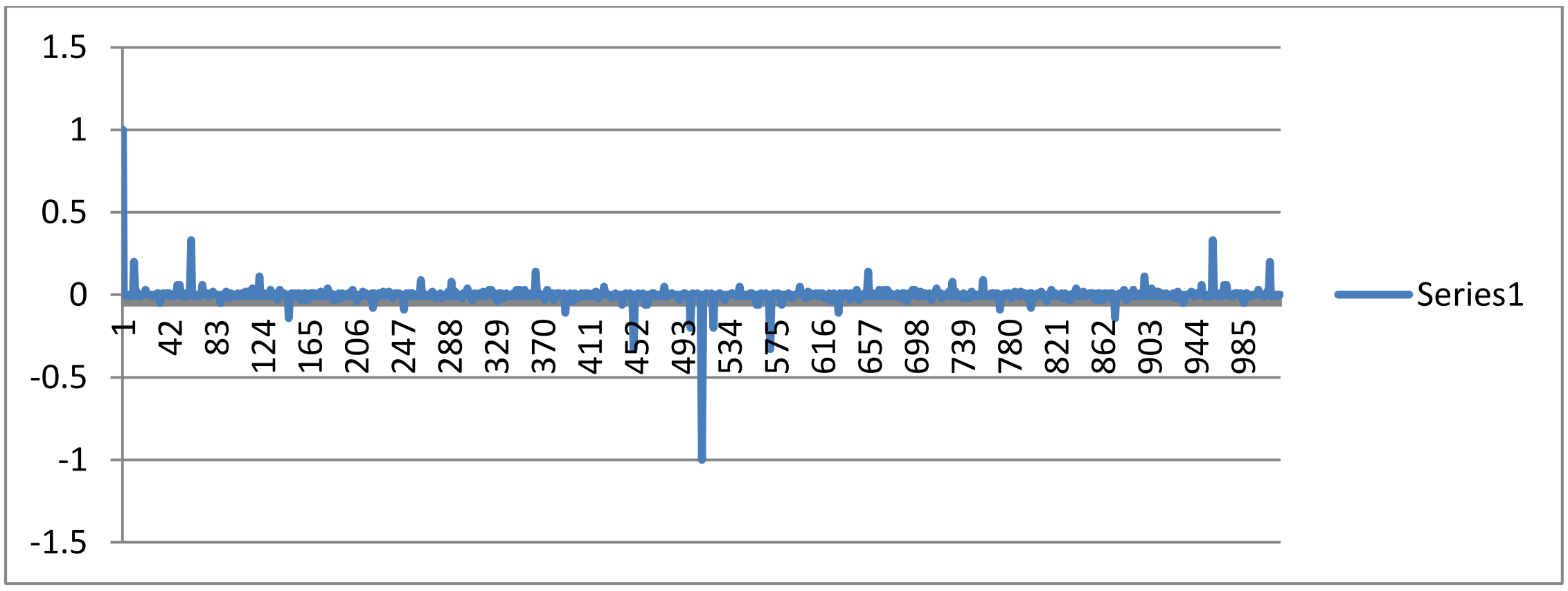

*Figure 6 Autocorrelation graph for input D- sequence*

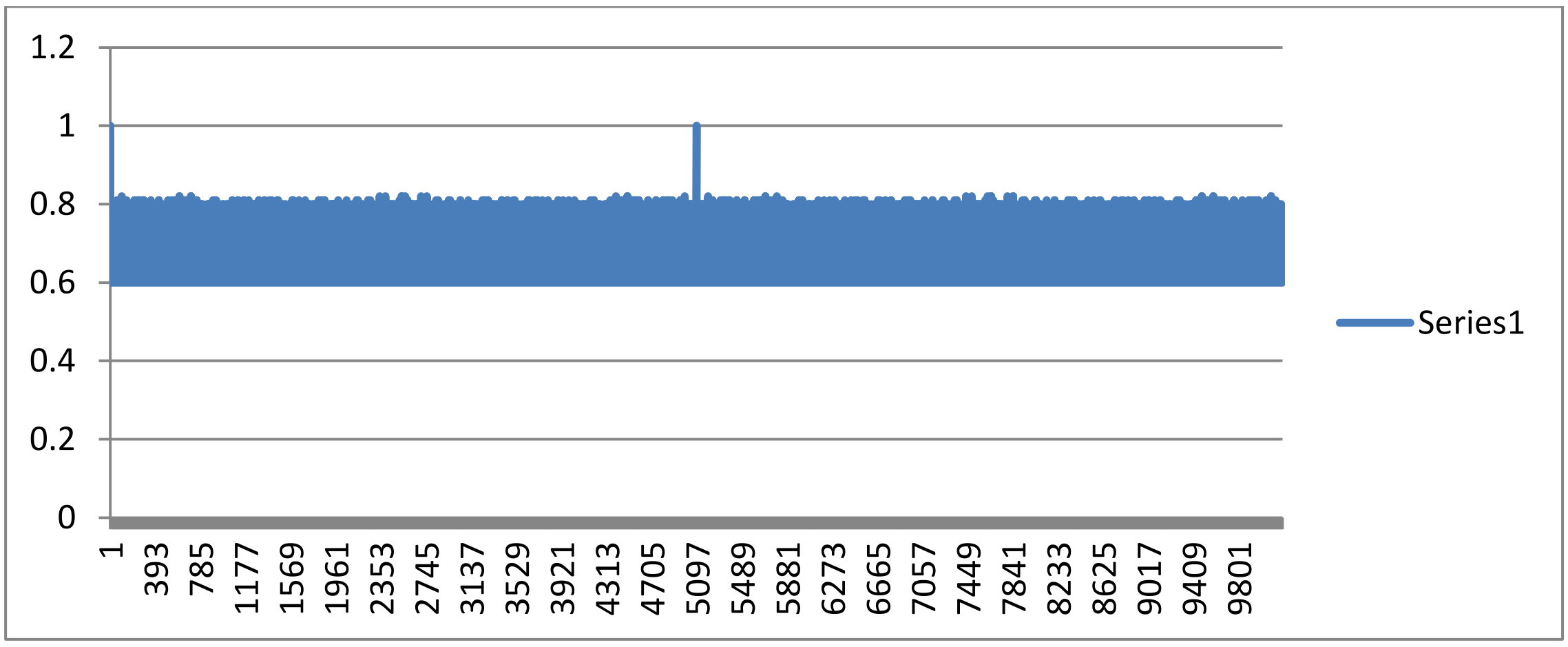

*Figure 7 Autocorrelation graph for the sequence generated by Convolutional encoder*

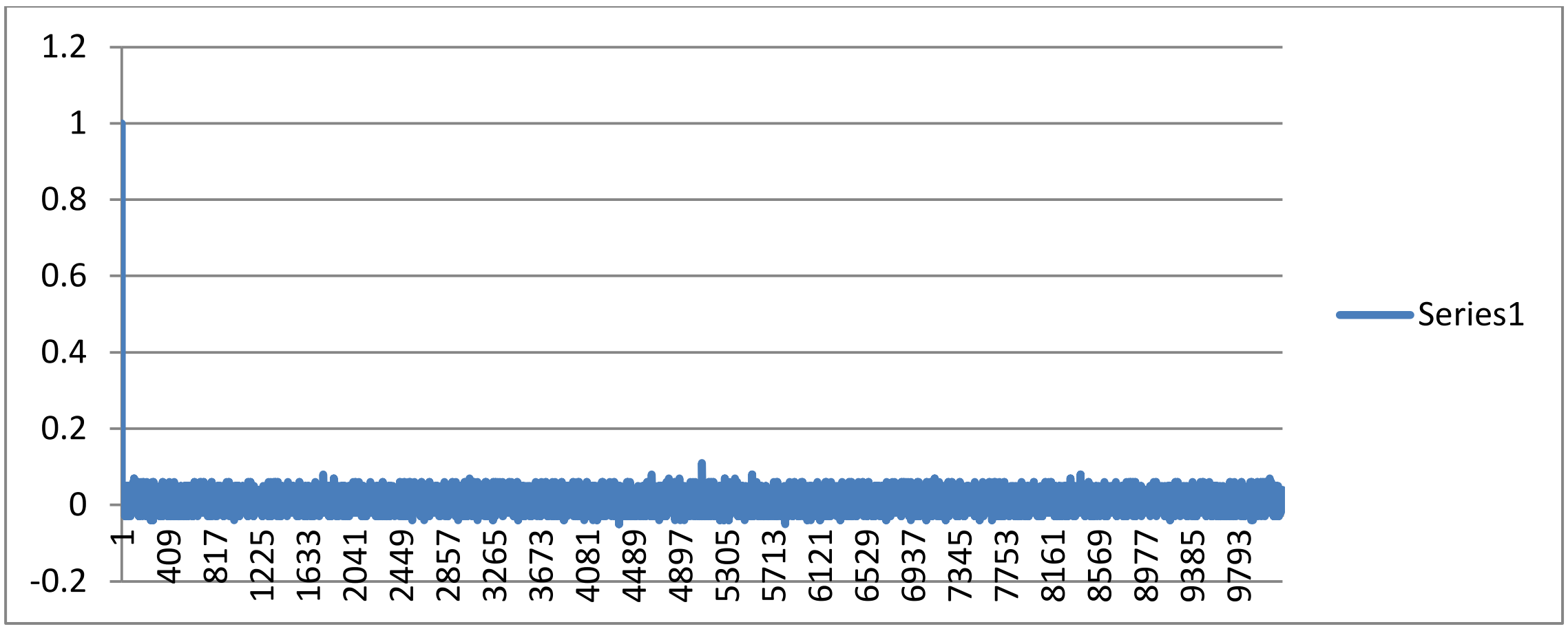

*Figure 8 Autocorrelation graph for the sequence generated by probability function*

**Example 3:** *Input type: D-sequence of size 353 and period is 87*

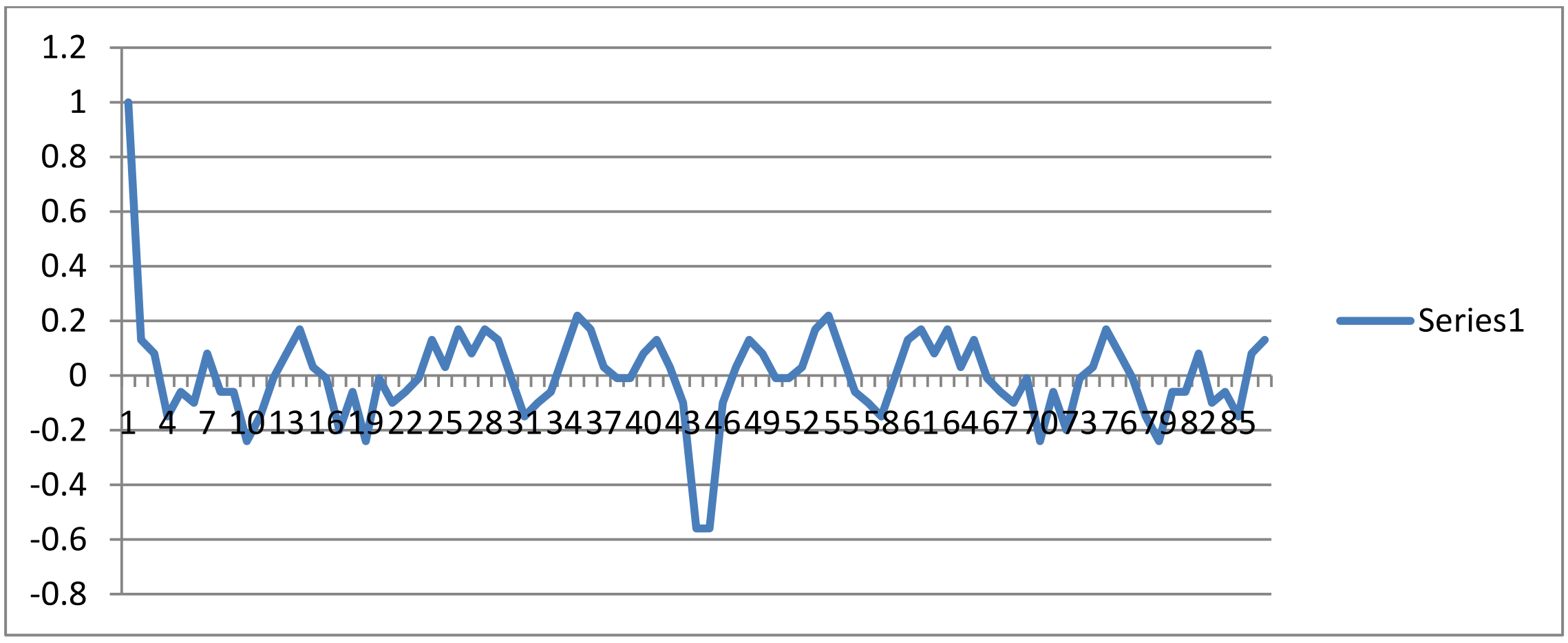

*Figure 9 Autocorrelation graph for input d- sequence*

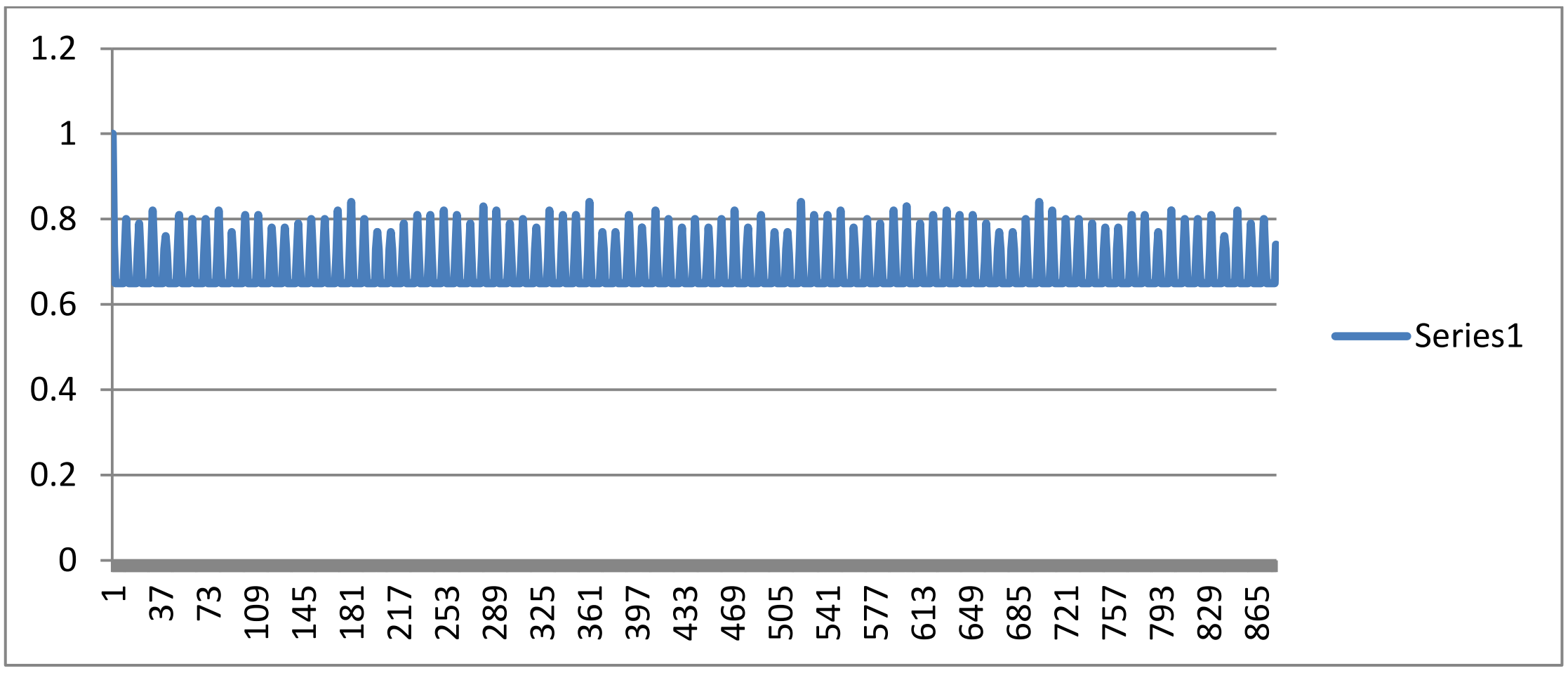

*Figure 10 Autocorrelation graph for the sequence generated by Convolutional encoder*

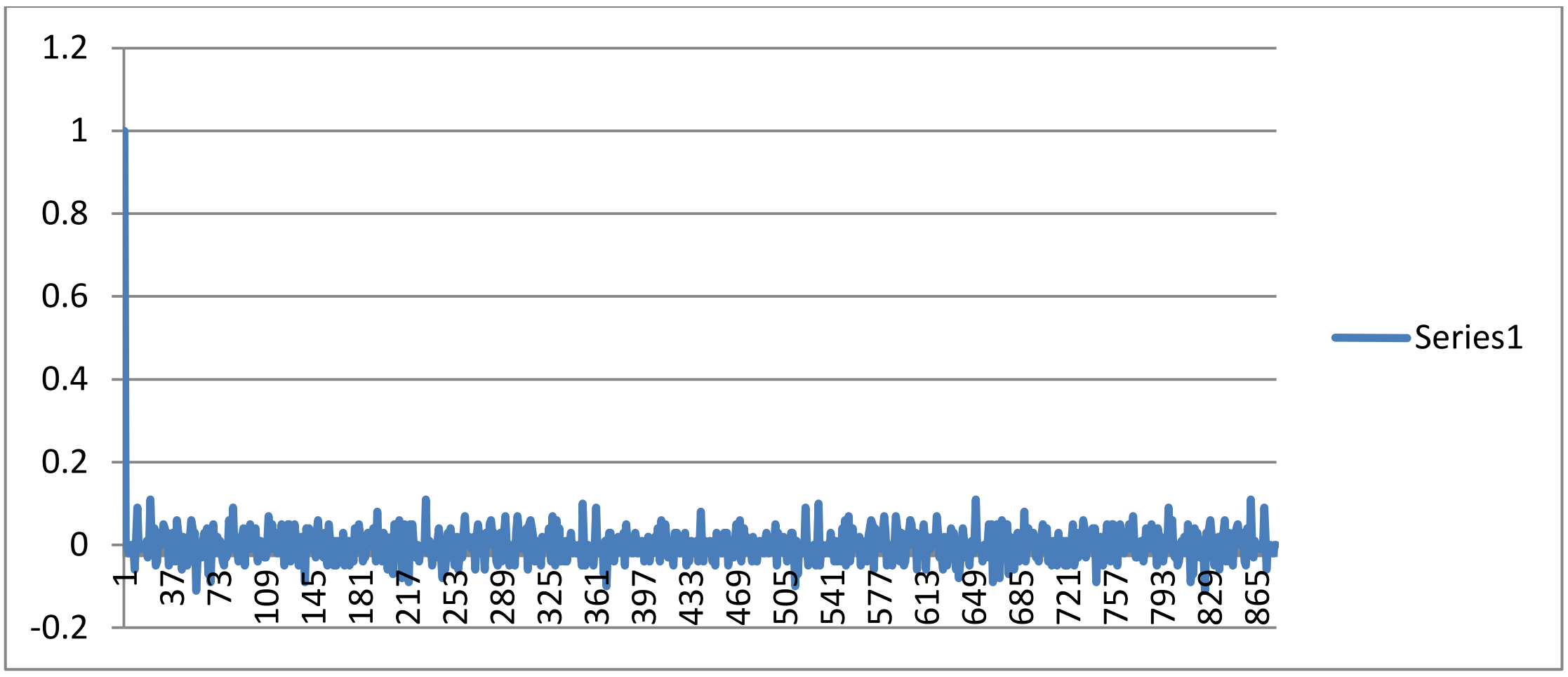

*Figure 11 Autocorrelation graph for the sequence generated by probability function*

**Example 4:** *Input Sequence all zeros*

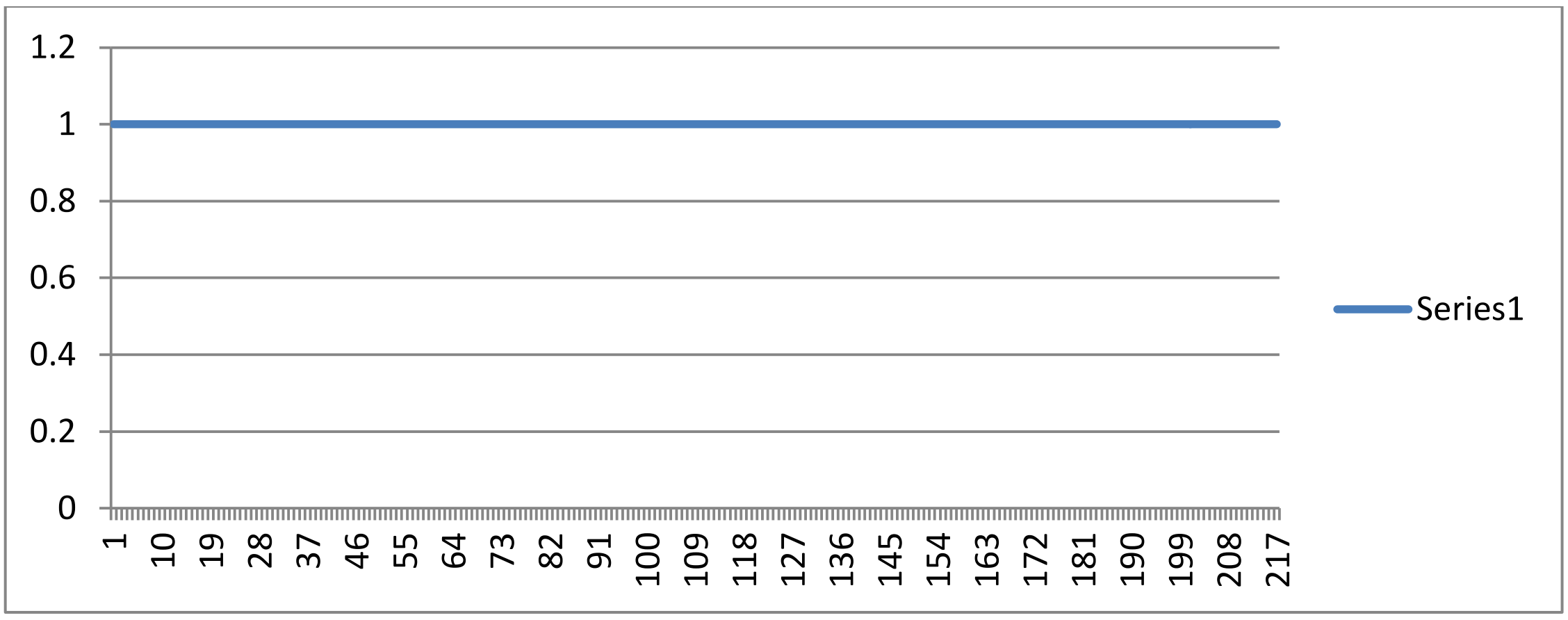

*Figure 12 Autocorrelation for input sequence*

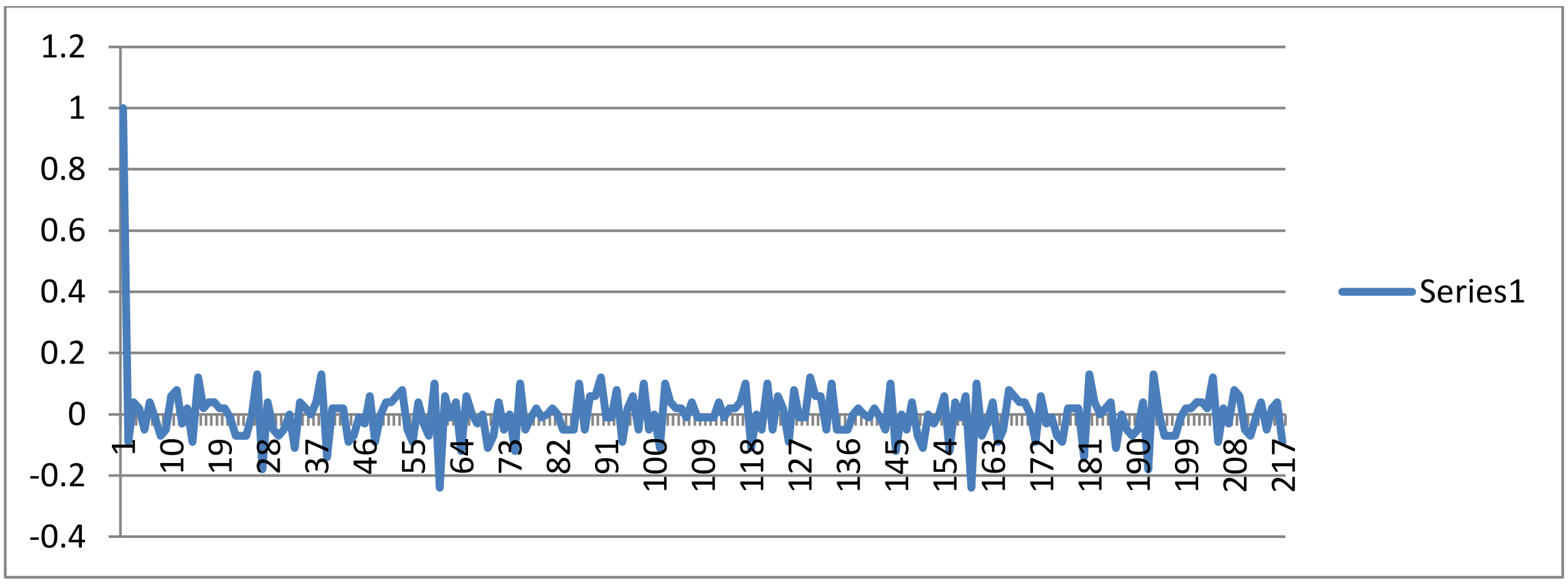

*Figure 13 Autocorrelation graph for output sequence obtained by probability function*

We see that in some of the examples the correlations in the initial sequence carry through to the output of the convolutional coder. But in the second step of introducing the probability mapping function, these correlations go away.

VII BLOCK WISE RANDOMIZATION

To improve the randomness of a sequence further, we have divided the sequence into equal size blocks and applied the same probability randomization technique on each block separately. The performance measures are shown below. We consider an extreme case of a sequence that is comprised of only 1s.
Example 1: Sequence containing 1100 zeros

Table 1 Block size versus Peak value of the autocorrelation graph

| Index of X-axis | Block size | Peak value |
|---|---|---|
| 1 | 550 | 0.11 |
| 2 | 275 | .10 |
| 3 | 220 | .08 |
| 4 | 110 | .09 |
| 5 | 55 | .09 |
| 6 | 50 | .09 |
| 7 | 44 | .1 |
| 8 | 25 | .08 |
| 9 | 22 | .09 |
| 10 | 20 | .08 |
| 11 | 11 | .09 |
| 12 | 10 | .08 |
| 13 | 5 | .15 |
| 14 | 4 | .16 |
| 15 | 2 | .2 |

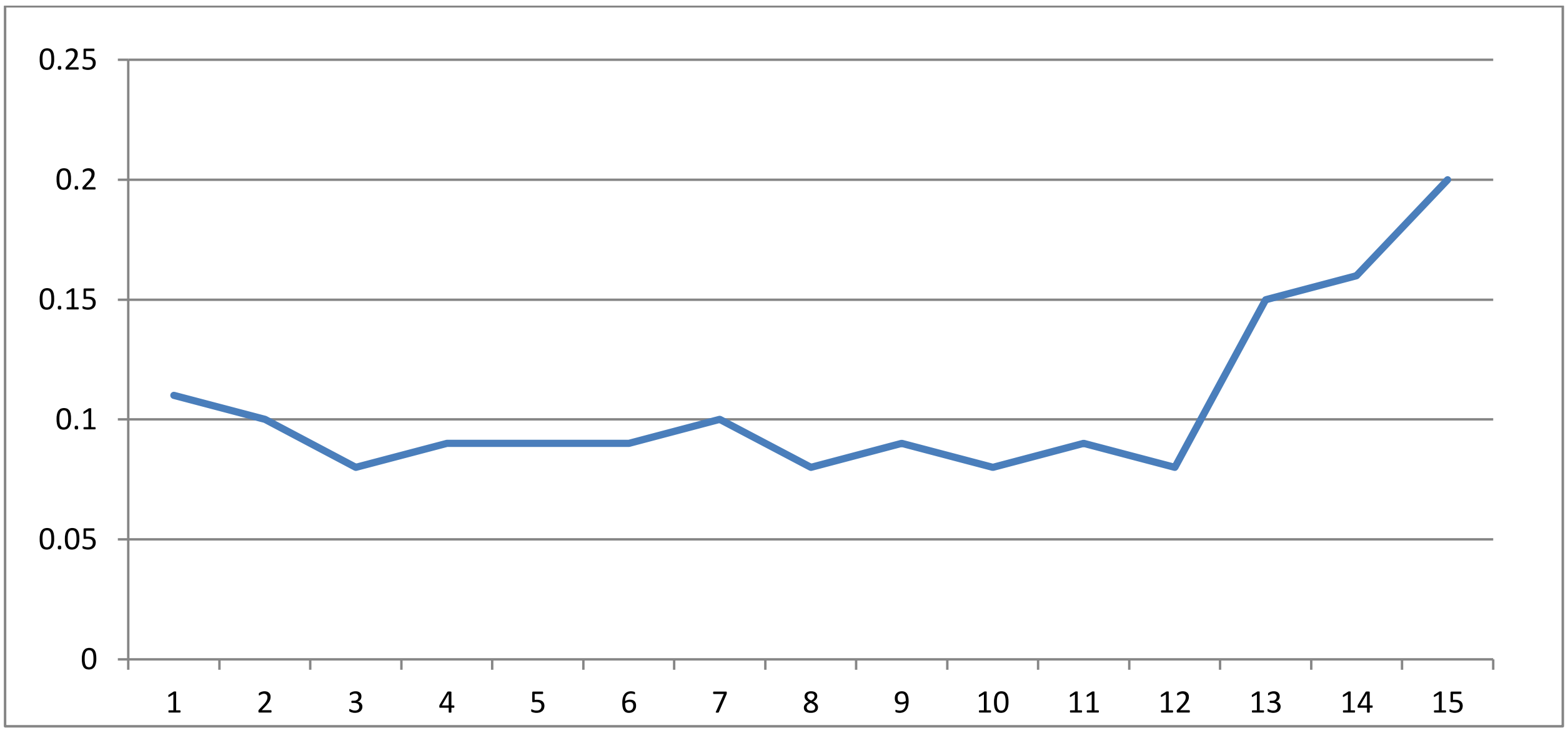

Figure 14 Graph showing the relationship between block size and randomness of the sequence generated.

The table below clearly shows that as the number of blocks is increased, randomness of the sequence increases till certain number of blocks. Later the randomness again decreases as the number of blocks is increased further. This is because if the block size is too small though we make number of zeros and ones equal it will not make the entire sequence more random.

Example 2: 353-D-sequence

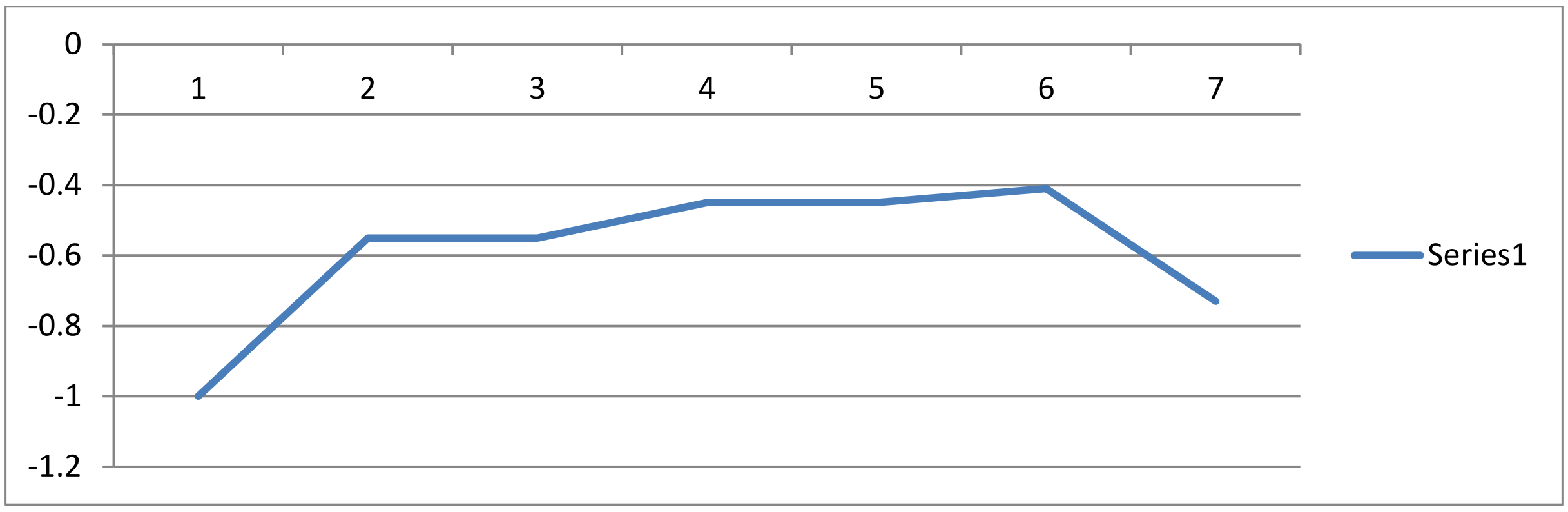

Figure 15 Graph showing the relationship between block size and randomness of the sequence generated.

The Table below shows Block size versus Peak value of the autocorrelation graph

| | Block Size | Peak value |
|---|---|---|
| 1 | Without blocks | -1 |
| 2 | 2 | -0.73 |
| 3 | 4 | -0.41 |
| 4 | 8 | -0.45 |
| 5 | 11 | -0.45 |
| 6 | 22 | -0.55 |
| 7 | 44 | -0.55 |

From the above two examples we can conclude that the block size should not be too small or too large. We need to select the block size in such a way that it should fall in the mid-range of maximum and minimum block sizes of the sequence.

## VIII LIMITATIONS AND FUTURE WORK

The limitation of this paper is that the method is need of generalization so that its performance is not affected by the nature of the input. As it stands, while converting 1s to 0s blocks of 0s remains unchanged and while converting 0s to 1s blocks of 1s remains unchanged.

Example: Take the sequence 111111110000. Here, the number of 1s is eight and the number of 0s is four. Now we need to convert four 1s to 0s. Java random function chooses four random 1s to make them to 0s while the block of 0s remains unchanged which eventually diminishes the randomness (lack of diffusion) of the sequence.

## IX CONCLUSION

We have shown that the proposed system which makes use of convolutional codes and probability functions provides a method of generating random output sequence. The results show that the randomization obtained is very good and the sequences obtained are cryptographically strong. A system based on these ideas can also be used to generate hashes.